# Fractal Polariton Topological Insulator

## KHALIL SABOUR[1,*] AND YAROSLAV V. KARTASHOV[2]

[1]*Moscow Institute of Physics and Technology, Institutsky lane 9, Dolgoprudny, Moscow region, 141701, Russia*
[2]*Institute of Spectroscopy, Russian Academy of Sciences, 108840, Troitsk, Moscow, Russia*
*Corresponding author: Khalilsabor1998@gmail.com*

**We introduce higher-order polariton topological insulator (HOTI) realized with fractal array of microcavity pillars arranged into Sierpiński gasket-like geometry. This system exhibiting self-similarity in different generations, can support localized modes either in the external or multiple internal corners depending on the controllable distortion introduced into first-generation structure. Nonlinear corner modes can be selectively excited in this dissipative system using resonant optical pumping. Strong polariton–polariton interactions lead to tilt of the resonance curves and bistability as the pump amplitude increases, allowing control over corner state profiles. Linear stability analysis illustrates dynamical stability of such states. Our results show how nontrivial topology manifests itself in self-similar, *aperiodic* structures and indicate that fractal geometry may bring qualitatively new localization scenarios in polariton HOTIs.**

Fractal structures are widespread in nature [1]. Their most representative feature is self-similarity, when each subsequent generation of a fractal is constructed using several copies of previous-generation structures allowing to reproduce the entire hierarchy. Their non-integer dimensionality is described by the Hausdorff dimension $d_\text{f}=\log_l(\text{m})$, with $m$ is the number of previous-generation elements required for the construction of next-generation fractal, while $l$ is the edge-scaling factor. Thus, the famous Sierpiński gasket fractal has Hausdorff dimension $d_\text{f}=\log_2(3)$. Such aperiodic structures find applications in solid-state physics, acoustics, and photonics [2], where their unique geometry enables novel localization, transport, and spectral features inaccessible in uniform or periodic systems.

Fractals can be used in construction of topological systems where nontrivial topology produces localized boundary states [3–5]. Their internal structure is especially promising for realizing higher-order topological insulators (HOTIs) [6,7], that host at the boundaries of $d$-dimensional structure topological states with effective dimensionality $d-q$, where $q>1$ is the order of HOTI. Fractal topological insulators have been studied in electronic, photonic, and acoustic platforms [8–16], mostly in linear conservative system. Nonlinear fractal HOTIs were also predicted [12,13]. A unique aspect of fractals is that their real-space self-similarity alone can induce topological features, without engineered couplings typically required for topological phases [17]. Unlike periodic lattices that require fine-tuned couplings [17], fractals such as the Sierpiński gasket may naturally support topological phases through geometry alone, making them compelling for geometry-induced topological phenomena.

Among dissipative and nonlinear systems, polariton condensates in microcavities offer a versatile platform for realization of fractal HOTIs. Topological polariton insulators support edge states and solitons [18–22], bistable edge states [23–25], and allow observation of topological transitions in 1D chains [26–28]. Recently, polariton microcavities were proposed as platform for realization of dissipative HOTIs [29–31]. Two-dimensional polariton HOTIs based on Su–Schrieffer–Heeger lattices have been realized [32,33], though not yet in aperiodic fractal structures.

In this Letter, we propose the first realization of a dissipative *fractal* HOTI for polaritons implemented using *aperiodic* array of the microcavity pillars in the Sierpiński gasket geometry. The unique property of the system stemming from its fractality is that it remains in topological phase practically for any shift of the pillars in first-generation structure used to construct all hierarchy with structure of corner states qualitatively changing for different directions of the shift. The key advantage of the proposed system is the possibility to use uniform *resonant* optical pump with tunable frequency for qualitative modification of the excited polariton patterns and *selective* excitation of strongly localized and topologically protected corner states, affording remarkable control over spatial emission from microcavity (from extended to localized in outer or inner corners). Repulsive polariton–polariton interactions lead to tilted resonance curves and bistability effects, offering additional control over shapes of nonlinear corner modes of topological origin, that can be *dynamically stable* in a broad range of frequencies. This offers advantage over conventional topological polariton lasers, where lasing becomes strongly multimode slightly above the condensation threshold, and is promising for the enhancement of any nonlinear processes involving corner modes and design of compact radiation sources.

We consider the evolution of polariton condensate in the array of microcavity pillars, governed by the dimensionless Gross-Pitaevskii equation for the polariton wavefunction $\psi$ [4,18,28]:

$$i\frac{\partial \psi}{\partial t}=-\frac{1}{2}\Delta\psi-\mathcal{R}(x,y)\psi+|\psi|^2\psi-i\alpha\psi+\mathcal{P}_0 e^{-i\varepsilon t}. \qquad (1)$$

Here $\Delta=\partial^2/\partial x^2+\partial^2/\partial y^2$; the coordinates $(x,y)$ are scaled to the characteristic length $\ell$ setting energy scale $\varepsilon_0=\hbar^2/m_\text{eff}\ell^2$, $m_\text{eff}$ is the effective polariton mass; the evolution time $t$ is normalized to $\hbar\varepsilon_0^{-1}$; $\varepsilon$ is the pump frequency detuning from the bottom of the lower polariton branch normalized to $\hbar^{-1}\varepsilon_0$; $\alpha=\hbar/\tau\varepsilon_0$ is the loss coefficient, $\tau$ is the polariton lifetime; $\psi=(g/\varepsilon_0)^{1/2}\Psi$ is the dimensionless wavefunction; $g$ is the interaction constant; $\mathcal{R}(x,y)=p\sum_{m,n}e^{[-(x-x_{m,n})^2-(y-y_{m,n})^2]/\sigma^2}$ is the potential energy landscape scaled to $\varepsilon_0$ and composed from wells with half-width $\sigma$, depth $p$, spacing $a$ between next-nearest wells; $x_{m,n}, y_{m,n}$ are the coordinates of micropillars. Nonlinear term accounts for repulsion between polaritons; and $\mathcal{P}_0 e^{-i\varepsilon t}$ is the external resonant pump. Setting $\ell=1\ \mu\text{m}$, $m_{\text{eff}}=5\times10^{-5}m_e$, one obtains the characteristic energy $\varepsilon_0 \approx 1.52$ meV and time $\hbar\varepsilon_0^{-1} \approx 0.43$ ps. We set $p=6$, $\sigma=0.9$, and $\alpha=0.01$. This corresponds to potential energy landscape with $\sim 9$ meV wells of width $\sim 2\ \mu\text{m}$ with next-nearest spacing of 6 $\mu$m and polariton lifetime $\sim 40$ ps typical for experiments with high-Q microcavities [4,23]. Resonant excitation of polaritons in microcavities controlled by laser frequency detuning from the bottom of lower polariton branch is reported in [28].

Sierpiński gasket micropillar array $\mathcal{R}$ is constructed recursively from 3 triangular units of previous generation. The first-generation triangle $G_1$, outlined by the blue dashed line in Fig. 1(a), serves as the simplest building block. The second-generation structure $G_2$ is formed by combining three copies of $G_1$, with each pair sharing a single site, as indicated by the cyan dashed line. Repeating the procedure, one gets third-generation structure $G_3$, shown in Fig. 1(a). Each generation $G_k$ contains $T_k = 3^{k+1} - (3^k-3)/2$ micropillars, where $k$ is the generation index. Due to the method of construction, such fractal arrays possess multiple holes, inner corners and edges. To realize a fractal HOTI, we introduce a controlled geometric distortion parameter $r$. The distortion is introduced into $G_1$ by shifting neighboring pillars along the triangle edges in the opposite directions, while keeping the spacing $a=6$ between next-nearest-neighbors fixed. Distorted Sierpiński gasket arrays at $r=0.38a$ and $r=0.62a$ are presented in Fig. 1(b) and 1(c), respectively, while undistorted array from Fig. 1(a) corresponds to $r=0.50a$.

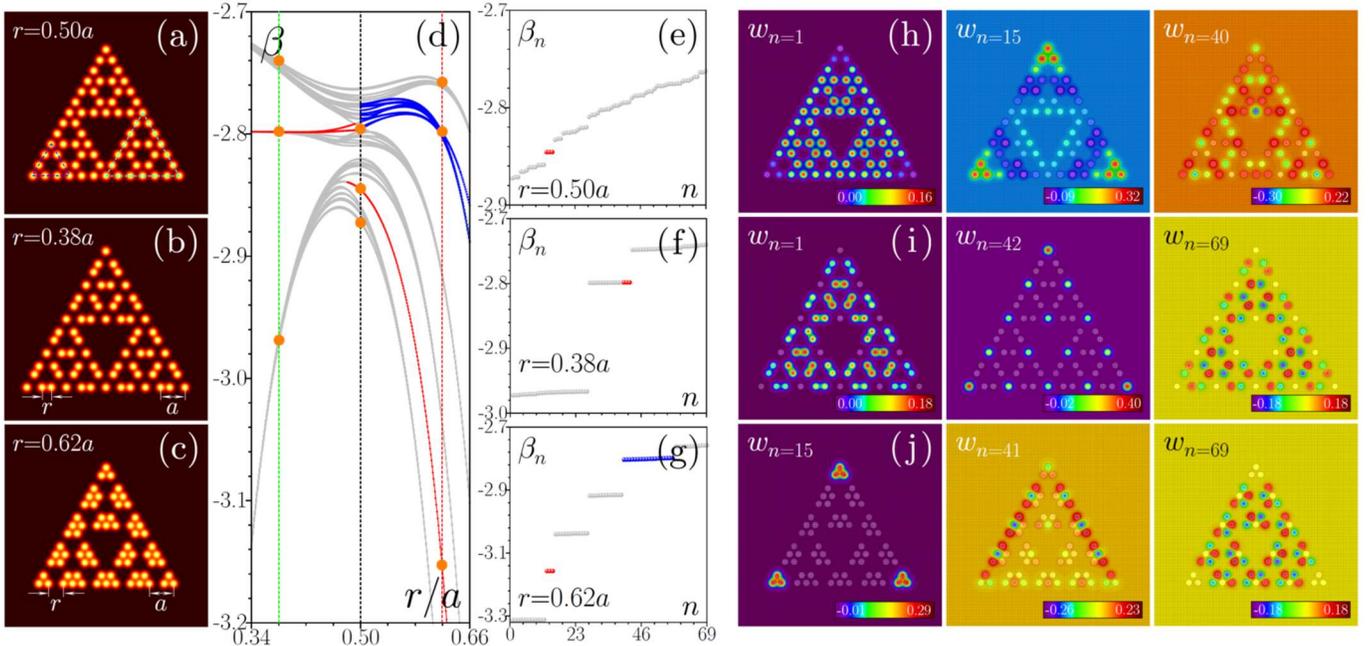

Fig. 1. (a)-(c) Fractal arrays of microcavity pillars of third generation $G_3$ for different distortion parameters $r$; first- $G_1$ and second-generation $G_2$ elements are shown with blue and cyan dashed lines. (d) Linear eigenvalue spectrum versus $r$: grey dots–delocalized states, red–corner states, blue–edge states. Vertical dashed lines correspond to $r=0.38a$, $r=0.50a$, and $r=0.62a$; orange dots correspond to representative eigenmodes in (h)-(j). (e)-(g) Eigenvalues $\beta_n$ of all modes in ascending order versus mode index $n$ at marked $r$ values in (d). Profiles of eigenmodes $w_n$ at $r=0.50a$ (h), $r=0.38a$ (i), and $r=0.62a$ (j). Here and in all figures below $p=6$, $\sigma=0.9$, and $a=6$.

Shift of micropillars $r$ strongly affects linear spectrum of the array. To obtain it, we omit nonlinearity, losses, and pump from Eq. (1) and calculate linear eigenmodes using the ansatz $\psi=w(x,y)e^{-i\beta t}$, which yields a linear eigenvalue problem $\beta w = -(1/2)\Delta w - \mathcal{R}w$. In the absence of losses, the eigenvalues $\beta$ and eigenmodes $w$ are real. All lowest eigenvalues at $\beta<0$ (corresponding to modes localized within the array region) are plotted in Fig. 1(d) as a function of scaled shift $r/a$. In these figures, grey circles correspond to extended states, red ones to localized corner modes, while blue circles correspond to edge modes. The complete set of eigenvalues $\beta$, sorted in ascending order by mode index $n$, is shown in Fig. 1(e)-1(g) for $r=0.50a$, $r=0.38a$, and $r=0.62a$. The remarkable feature of fractal array is that corner modes in it can appear both at $r>0.5a$ [in this case, they emerge in outer corners, see mode $w_{n=15}$ in Fig. 1(j)] and at $r<0.5a$ [in this case, they emerge also in all *inner* corners, see hybrid mode $w_{n=42}$ in Fig. 1(i)]. Fractal array is aperiodic structure; hence its topological properties can be characterized using real-space polarization index $\mathcal{P}$ (see [13,16] for its definition). Its calculation for corner states in the bandgap [red lines in Fig. 1(d)] for a given $r$ yields quantized value $\mathcal{P}=1/2$ for corner states existing at both $r>0.5a$ and $r<0.5a$ that hints on the fact that fractal array is topological for nearly all values of $r$. Similar conclusions are drawn in [17]. At the same time, $\mathcal{P}=0$ for bulk modes with lowest eigenvalue $\beta$. Being states of topological origin, corner modes are not destroyed by weak disorder that does not close the gap, where such states reside. Examples of other eigenmodes, corresponding to orange dots in Fig. 1(d), including extended and edge ones are presented in Fig. 1(h)-1(j). Notice that corner state $w_{n=15}$ in Fig. 1(j) at $r>0.5a$ can be considered effectively zero-dimensional, as in usual HOTIs, as number of pillars, where the amplitude $|\psi|$ is large, $C_k = 9$ is constant and the ratio $C_k/T_k \to 0$ with increase of generation index $k$. In contrast, the state $w_{n=42}$ from Fig. 1(i) at $r < 0.5a$ has large amplitude on $C_k = (3^k+3)/2$ pillars, so that $C_k/T_k \to 0.2$ as $k \to \infty$, hinting on the fact that the effective dimensionality of this hybrid mode coincides with that of the fractal array.

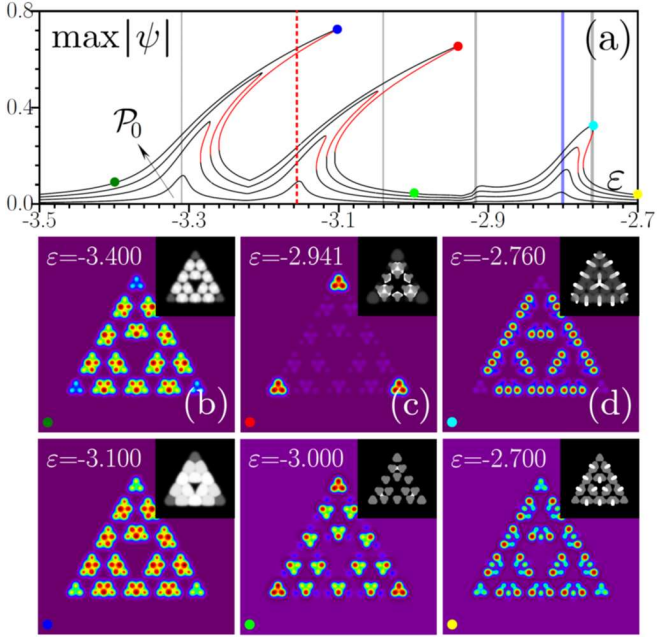

Fig. 2. (a) Peak amplitude $\max|\psi|$ vs detuning $\varepsilon$ in fractal HOTI with $r = 0.62a$ for pump amplitudes, $\mathcal{P}_0 = 0.0005$, $0.0015$, $0.0025$ and $0.0035$ (arrow indicates increasing $\mathcal{P}_0$). Grey–extended states, blue region–edge band, dashed red—corner state eigenvalue. Examples of modulus $|\psi|$ and phase $\arg(\psi)$ (insets) in condensates corresponding to color dots in (a) around resonances with extended (b), corner (c) and edge (d) states.

Resonant optical pumping in the presence of losses and polariton-polariton interactions enables the selective excitation of different states in fractal HOTI, provided that the pump energy detuning $\varepsilon$ matches the eigenvalue $\beta$ of the targeted state. Figure 2(a) illustrates the dependence of the peak amplitude $a = \max|\psi|$ of extended, corner, and edge states, excited by a uniform pump $\mathcal{P}_0$, on pump frequency detuning $\varepsilon$ for $r = 0.62a$. The states were computed from Eq. (1) in the form $\psi = w(x,y)e^{-i\varepsilon t}$ using the Newton iteration method. The overlap of pump with corresponding linear modes determines the efficiency of excitation of the nonlinear states. In Fig. 2(a) at $r = 0.62a$ it is considerable for extended states from lowest gray band, for topological corner mode from Fig. 1(j), whose eigenvalue is shown with dashed red line, and with edge states from blue band. While at low pump amplitudes the tip of the resonance coincides with linear eigenvalue of corresponding states, with increase of $\mathcal{P}_0$ resonances tilt due to nonlinearity leading to bistability. Profiles $|\psi|$ and phase distributions ($\arg(\psi)$, insets) in states corresponding to colored dots in resonance curves are shown in Fig. 2(b)-2(d). Extended, corner, and edge states feature qualitatively different phase distributions. The contributions from different types of states into excited nonlinear solution notably changes with $\varepsilon$, with most efficient excitation of state of given type (corner or edge) occurring in the tips of resonances that notably shift with increase of $\mathcal{P}_0$, while dots between resonances correspond to delocalized states that represent a mixture of states of different types. In this way nonlinearity in this system admitting different types of modes allows to tune the frequencies corresponding to the most efficient excitation of, e.g., corner states (already after the array is fabricated and its linear spectrum is set), that is important from the practical point of view.

In fractal HOTI with $r = 0.38a$ uniform pump $\mathcal{P}_0$ efficiently excites extended states from the lowest band and a different type of topological state appearing in all corners (including internal ones) of the structure, see resonance curves in Fig. 3(a). Again, increasing pump amplitude leads to suppression of the background inside the array in the tip of the resonance leading to enhanced localization in inner/outer corners, see Fig. 3(c), the state corresponding to the red dot. Thus, the combination of polariton-polariton interactions and resonant pump allows highly selective and tunable excitation of topological states with very different internal structure.

To illustrate stability of nonlinear corner states we employ linear stability analysis by writing perturbed states in the form $\psi = [w(x,y) + u(x,y)e^{\lambda t} + v^*(x,y)e^{\lambda^* t}]e^{-i\varepsilon t}$, where $u, v \ll w$, and $\lambda = \lambda_{\mathrm{re}} + i\lambda_{\mathrm{im}}$ is the complex perturbation growth rate. Linearization of Eq. (1) around $w$ yields in this case the eigenproblem:

$$\lambda u = i[+(1/2)\Delta u + \mathcal{R}u + i\alpha u + \varepsilon u - 2|w|^2 u - w^2 v],$$
$$\lambda v = i[-(1/2)\Delta v - \mathcal{R}v + i\alpha v - \varepsilon v + 2|w|^2 v + w^{*2} u]. \quad (2)$$

The state $w$ is stable when $\lambda_{\mathrm{re}} \leq 0$ for all possible perturbations and is unstable when $\lambda_{\mathrm{re}} > 0$ at least for one of them. Solution of (2) confirms stability of all lower and upper branches of the resonance curves for all types of excited states [black lines in Fig. 2(a) and 3(a)], even when resonances exhibit significant tilt, for both $r = 0.62a$ and $r = 0.38a$, and indicates on expected instability of all middle branches (red lines). Representative $\lambda_{\mathrm{re}}(\varepsilon)$ dependencies for shift $r = 0.62a$ are shown in Fig. 4(a).

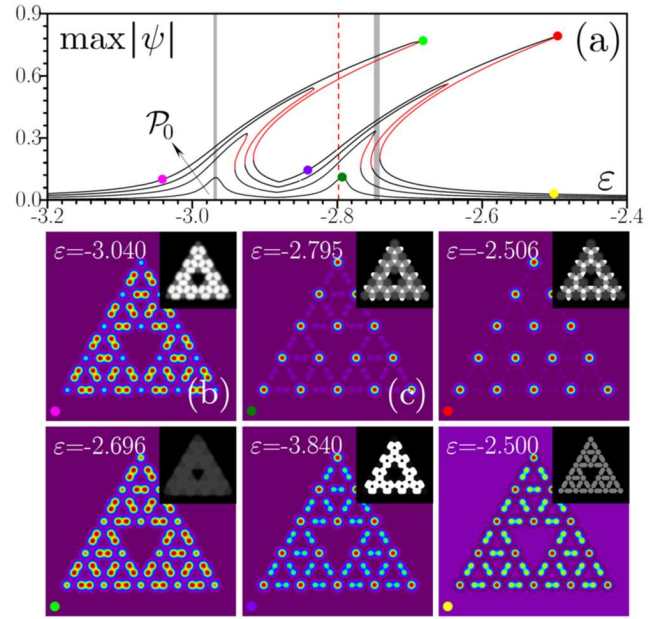

Fig. 3. The dependencies as in Fig. 2, but for $r = 0.38a$. Examples of modulus $|\psi|$ and phase $\arg(\psi)$ (insets) distributions of condensates are shown around resonances with extended (b) and corner states (c).

To confirm stability of different corner states beyond the perturbative approach, we modeled their evolution in Eq. (1) in the presence of broadband noise $\xi(x,y)$ (up to $5\%$ in amplitude) added into input states. Figure 4(b) illustrates typical stable evolution of corner states from upper branches of resonance at $r=0.62a$ and $r=0.38a$ (black lines) and decay of states from middle branches (red lines) for pump amplitude $\mathcal{P}_0=0.0035$. In the former case the perturbed corner states "clean up" from noise at typical times $\sim 5\alpha^{-1}$ that is consistent with results of the linear stability analysis and then keep their structure with time as long as pump is on [Fig. 4(c)]. The states from the middle branch "jump" via oscillations onto upper or lower branches [Fig. 4(d)].

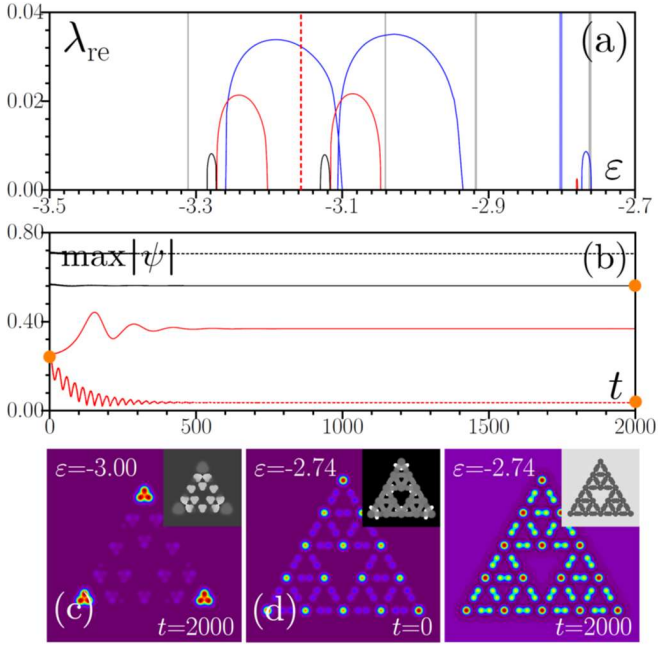

Fig. 4. (a) Maximal $\lambda_{\rm re}$ for unstable middle branches at $r=0.62a$ vs $\varepsilon$ for $\mathcal{P}_0=0.0015$ (black line), $0.0025$ (red), and $0.0035$ (blue). (b) Peak amplitude vs time for perturbed states: upper $(\varepsilon=-3.00)$, middle $(\varepsilon=-3.10)$ branches at $r=0.62a$ (solid curves), and upper $(\varepsilon=-2.56)$, middle $(\varepsilon=-2.74)$ branches at $r=0.38a$ (dashed curves). Amplitude dependence for upper branches—black; for middle ones—red. Examples of states from upper (c) and middle (d) branches at different $t$ corresponding to orange dots in (b).

In summary, we studied the properties of dissipative topological states in fractal polaritonic HOTI created on the Sierpiński gasket array of microcavity pillars. We have shown that these structures support two different types of corner modes of topological origin that can be excited using resonant external plane wave pump. Our results highlight the unique interplay between fractal geometry and nonlinearity in dissipative system.

**Funding:** This work was supported by the Russian Science Foundation (grant 24-12-00167) and partially by the research project FFUU-2024-0003 of the Institute of Spectroscopy of the Russian Academy of Sciences.

**Disclosures:** The authors declare no conflicts of interest.

**Data availability.** Data underlying the results presented in this paper may be obtained from the authors upon reasonable request.